\documentstyle[11pt,epsfig]{article}
\begin{document}
%\begin{document}
%\special{papersize=8.26in,11.69in}
%\textwidth15.0cm
%\textheight22.0cm
%\baselineskip1.0cm
%\setlength{\topmargin}{-1cm}
%\addtolength{\textheight}{1cm}
%\oddsidemargin+1.2cm
%\evensidemargin-1.2cm
\pagestyle{plain}

%%%%%%%%%%%%%%% insert actual file mydefs.sty %%%%%%%%%%%%%%%%%%%%%%
%
\newcommand{\be}{\begin{equation}}
\newcommand{\ee}{\end{equation}\noindent}
\newcommand{\bear}{\begin{eqnarray}}
\newcommand{\ear}{\end{eqnarray}\noindent}
\newcommand{\no}{\noindent}
\date{}
\renewcommand{\theequation}{\arabic{section}.\arabic{equation}}
\renewcommand{\arraystretch}{2.5}
\newcommand{\GeV}{\mbox{GeV}}
\newcommand{\cL}{\cal L}
\newcommand{\D}{\cal D}
\newcommand{\Dhalf}{{D\over 2}}
\newcommand{\Det}{{\rm Det}}
\newcommand{\PP}{\cal P}
\newcommand{\G}{{\cal G}}
\def\GBd12{{\dot G}_{B12}}
\def\R{1\!\!{\rm R}}
\def\Eins{\mathord{1\hskip -1.5pt
\vrule width .5pt height 7.75pt depth -.2pt \hskip -1.2pt
\vrule width 2.5pt height .3pt depth -.05pt \hskip 1.5pt}}
\newcommand{\symb}{\mbox{symb}}
\renewcommand{\arraystretch}{2.5}
\newcommand{\slD}{\raise.15ex\hbox{$/$}\kern-.57em\hbox{$D$}}
\newcommand{\slpartial}{\raise.15ex\hbox{$/$}\kern-.57em\hbox{$\partial$}}
\newcommand{\slG}{{{\dot G}\!\!\!\! \raise.15ex\hbox {/}}}
\newcommand{\Gd}{{\dot G}}
\newcommand{\Gund}{{\underline{\dot G}}}
\def\np{n_{+}}
\def\nm{n_{-}}
\def\Np{N_{+}}
\def\Nm{N_{-}}
\def\PITD{{(4\pi T)}^{-{D\over 2}}}
\def\non{\nonumber}
\def\beqn*{\begin{eqnarray*}}
\def\eqn*{\end{eqnarray*}}
\def\sy{\scriptscriptstyle}
\def\footstrut{\baselineskip 12pt}
\def\square{\kern1pt\vbox{\hrule height 1.2pt\hbox{\vrule width 1.2pt
   \hskip 3pt\vbox{\vskip 6pt}\hskip 3pt\vrule width 0.6pt}
   \hrule height 0.6pt}\kern1pt}
\def\slash#1{#1\!\!\!\raise.15ex\hbox {/}}
\def\dint#1{\int\!\!\!\!\!\int\limits_{\!\!#1}}
\def\bra#1{\langle #1 |}
\def\ket#1{| #1 \rangle}
\def\vev#1{\langle #1 \rangle}
\def\rightvac{\mid 0\rangle}
\def\leftvac{\langle 0\mid}
\def\dps{\displaystyle}
\def\sy{\scriptscriptstyle}
\def\half{{1\over 2}}
\def\third{{1\over3}}
\def\fourth{{1\over4}}
\def\fifth{{1\over5}}
\def\sixth{{1\over6}}
\def\seventh{{1\over7}}
\def\eigth{{1\over8}}
\def\ninth{{1\over9}}
\def\tenth{{1\over10}}
\def\pa{\partial}
\def\ddtau{{d\over d\tau}}
\def\ge{\hbox{\textfont1=\tame $\gamma_1$}}
\def\gz{\hbox{\textfont1=\tame $\gamma_2$}}
\def\gd{\hbox{\textfont1=\tame $\gamma_3$}}
\def\go{\hbox{\textfont1=\tamt $\gamma_1$}}
\def\gt{\hbox{\textfont1=\tamt $\gamma_2$}}
\def\gth{\hbox{\textfont1=\tamt $\gamma_3$}} 
\def\gf{\hbox{$\gamma_5\;$}}
\def\ie{\hbox{$\textstyle{\int_1}$}}
\def\iz{\hbox{$\textstyle{\int_2}$}}
\def\id{\hbox{$\textstyle{\int_3}$}}
\def\ldop{\hbox{$\lbrace\mskip -4.5mu\mid$}}
\def\rdop{\hbox{$\mid\mskip -4.3mu\rbrace$}}
\def\eps{\epsilon}
\def\epshalf{{\epsilon\over 2}}
\def\e{\mbox{e}}
\def\g{\mbox{g}}
\def\pa{\partial}
\def\kinb{{1\over 4}\dot x^2}
\def\kinf{{1\over 2}\psi\dot\psi}
\def\expk{{\rm exp}\biggl[\,\sum_{i<j=1}^4 G_{Bij}k_i\cdot k_j\biggr]}
\def\expp{{\rm exp}\biggl[\,\sum_{i<j=1}^4 G_{Bij}p_i\cdot p_j\biggr]}
\def\expshort{{\e}^{\half G_{Bij}k_i\cdot k_j}}
\def\expabb{{\e}^{(\cdot )}}
\def\epseps#1#2{\varepsilon_{#1}\cdot \varepsilon_{#2}}
\def\epsk#1#2{\varepsilon_{#1}\cdot k_{#2}}
\def\kk#1#2{k_{#1}\cdot k_{#2}}
\def\G#1#2{G_{B#1#2}}
\def\Gp#1#2{{\dot G_{B#1#2}}}
\def\GF#1#2{G_{F#1#2}}
\def\Dab{{(x_a-x_b)}}
\def\Dsq{{({(x_a-x_b)}^2)}}
\def\lag{( -\partial^2 + V)}
\def\4piTD{{(4\pi T)}^{-{D\over 2}}}
\def\4piT4{{(4\pi T)}^{-2}}
\def\TintmD{{\dps\int_{0}^{\infty}}{dT\over T}\,e^{-m^2T}
    {(4\pi T)}^{-{D\over 2}}}
\def\Tintm4{{\dps\int_{0}^{\infty}}{dT\over T}\,e^{-m^2T}
    {(4\pi T)}^{-2}}
\def\Tintm{{\dps\int_{0}^{\infty}}{dT\over T}\,e^{-m^2T}}
\def\Tint{{\dps\int_{0}^{\infty}}{dT\over T}}
\def\pint{{\dps\int}{dp_i\over {(2\pi)}^d}}
\def\Dx{\dps\int{\cal D}x}
\def\Dy{\dps\int{\cal D}y}
\def\Dpsi{\dps\int{\cal D}\psi}
\def\Tr{{\rm Tr}\,}
\def\tr{{\rm tr}\,}
\def\sumij{\sum_{i<j}}
\def\freeexp{{\rm e}^{-\int_0^Td\tau {1\over 4}\dot x^2}}
\def\arraystretch{2.5}
\def\Ge{\mbox{GeV}}
\def\dA{\partial^2}
\def\DA{\sqsubset\!\!\!\!\sqsupset}
\def\FFdual{F\cdot\tilde F}
%\font\tame = cmmi12 scaled\magstep1
%\font\tamt = cmmi12 scaled\magstep2
%-------------------------------------------------------------------------
% To change the LaTeX pagestyle
% example  DINA4 format DESY
%-----------------------------------------------------------------------
% uncomment any of these if you want numbering to respect the sections
%
% \renewcommand{\thesection}{\arabic{section}.}
% \renewcommand{\thesubsection}{\thesection\arabic{subsection}.}
% \renewcommand{\theequation}{{\protect\thesection\arabic{equation}}}
% \renewcommand{\thetable}{{\protect{\bf \thesection\arabic{table}}}}
% \renewcommand{\thetable}{{\protect{\thesection\arabic{table}}}}
% \renewcommand{\thefigure}{{\protect\bf\thesection\arabic{figure}}}
% \renewcommand{\thefigure}{{\protect\thesection\arabic{figure}}}
% \renewcommand{\textfraction}{0}
% \renewcommand{\topfraction}{1.00}
% \renewcommand{\bottomfraction}{1.00}
% \renewcommand{\baselinestretch}{1.1}
%-----------------------------------------------------------------------
% special symbols: real numbers, unit matrix, integers
%
\def\bbbr{{\rm I\!R}}
\def\bbbone{{\mathchoice {\rm 1\mskip-4mu l} {\rm 1\mskip-4mu l}
{\rm 1\mskip-4.5mu l} {\rm 1\mskip-5mu l}}}
\def\bbbz{{\mathchoice {\hbox{$\sf\textstyle Z\kern-0.4em Z$}}
{\hbox{$\sf\textstyle Z\kern-0.4em Z$}}
{\hbox{$\sf\scriptstyle Z\kern-0.3em Z$}}
{\hbox{$\sf\scriptscriptstyle Z\kern-0.2em Z$}}}}
%-------------------------------------------------------------------------
%-------------------------------------------------------------------------
%%%%%%%%%%%%%%%%%%%%%%%%%%%%%%%%%%%%%%%%%%%%%%%%%%%%%%%%%%%%%%%%%%%%%%%
%----------------------------------------------------------
% Title page
%\begin{document}
\pagestyle{empty}
\renewcommand{\thefootnote}{\fnsymbol{footnote}}
\hskip 9cm {\sl LAPTH-783/2000}
\vskip-.1pt
%\hskip 9cm {\sl HUB-EP-96/13}
%\vskip-.1pt
%\hskip 10cm hep-th/ 
\vskip .4cm
\begin{center}
{\Large\bf Vacuum Polarisation Tensors in Constant 
Electromagnetic Fields: Part II}
\vskip1.3cm

\vskip.5cm
 {\large Christian Schubert
%   \footnote{}
}
\\[1.5ex]
%{\it
%School of Natural Sciences, 
%Institute for Advanced Study\\
%Olden Lane, Princeton, NJ 08540, USA
%\medskip\\
%}
%and\\
{\it
Laboratoire d'Annecy-le-Vieux
de Physique Th{\'e}orique LAPTH\\
Chemin de Bellevue,
BP 110\\
F-74941 Annecy-le-Vieux CEDEX\\
FRANCE\\
schubert@lapp.in2p3.fr\\
}
\vskip.5cm

\vskip 2.5cm
 {\large \bf Abstract}
\end{center}
\begin{quotation}
\noindent
In the second part of this series we
apply
the ``string-inspired'' technique to the
calculation of one-loop
amplitudes involving both vectors and axialvectors,
as well as a general constant electromagnetic
background field.
The vector-axialvector two-point function
in a constant field is calculated explicitly.

\end{quotation}
\clearpage
\renewcommand{\thefootnote}{\protect\arabic{footnote}}
\pagestyle{plain}
%------------------------------------------------------

\setcounter{page}{1}
\setcounter{footnote}{0}

\section{Introduction: Standard Model
Processes in Constant Electromagnetic Fields}
\renewcommand{\theequation}{1.\arabic{equation}}
\setcounter{equation}{0}

In the first part of this series \cite{mevp}
we explained in detail how the
``string-inspired'' technique 
\cite{berkos,5glu,strassler,ss3} 
can be used
for the efficient calculation of one-loop
scalar/spinor QED $N$ - photon amplitudes
in a constant external field
\cite{ss1,mckshe,cadhdu,gussho,shaisultanov,rss,adlsch,shovkovy,ditsha}.
As an application we
calculated the scalar and spinor QED 
vacuum polarisation tensors in a constant
field. 

\begin{figure}[ht]
\begin{center}
\begin{picture}(0,0)%
\epsfig{file=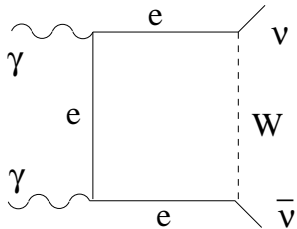}%
\end{picture}%
\setlength{\unitlength}{0.00047500in}%
\begingroup\makeatletter\ifx\SetFigFont\undefined
% extract first six characters in \fmtname
\def\x#1#2#3#4#5#6#7\relax{\def\x{#1#2#3#4#5#6}}%
\expandafter\x\fmtname xxxxxx\relax \def\y{splain}%
\ifx\x\y   % LaTeX or SliTeX?
\gdef\SetFigFont#1#2#3{%
  \ifnum #1<17\tiny\else \ifnum #1<20\small\else
  \ifnum #1<24\normalsize\else \ifnum #1<29\large\else
  \ifnum #1<34\Large\else \ifnum #1<41\LARGE\else
     \huge\fi\fi\fi\fi\fi\fi
  \csname #3\endcsname}%
\else
\gdef\SetFigFont#1#2#3{\begingroup
  \count@#1\relax \ifnum 25<\count@\count@25\fi
  \def\x{\endgroup\@setsize\SetFigFont{#2pt}}%
  \expandafter\x
    \csname \romannumeral\the\count@ pt\expandafter\endcsname
    \csname @\romannumeral\the\count@ pt\endcsname
  \csname #3\endcsname}%
\fi
\fi\endgroup
\begin{picture}(3834,2736)(120,-1050)
\end{picture}
\caption{\label{ggnnbar}
Diagram contributing to $\gamma\gamma\to\nu\bar\nu$
.}
\end{center}
\end{figure}

In the present sequel, we extend this analysis to the case of
mixed vector -- axialvector amplitudes in a constant electromagnetic
field. Amplitudes of this type have, in recent years, been much
investigated in connection with
photon-neutrino processes.
In the standard model, photon-neutrino interactions appear
at the one-loop level. 
In vacuum, a typical example would be the diagram
shown in fig. \ref{ggnnbar}, contributing to the process
$\gamma\gamma\to \nu\bar\nu$.

The $2\to 2$ processes
$\gamma\gamma\to\nu\bar\nu$, $\gamma\nu\to\gamma\nu$
and $\nu\bar\nu\to\gamma\gamma$ were considered
already before the advent of the standard model
using the Fermi theory \cite{chimor}.
However in the Fermi limit they vanish
due to the Landau-Yang theorem, as
was noted by Gell-Mann \cite{gellmann}
(for massless neutrinos, and with both photons on-shell).
In the standard model this suppression manifests itself
by factors of $\omega\over M_W$, where $\omega$ is the
center-of-mass energy and $M_W$ the $W$ boson mass.

There is no such suppression for processes involving two
neutrinos and more than two photons, which in vacuum are therefore
more important at low energies than the four-leg processes
\cite{hiesha,dicrep,dikare,abmapi,matias}.
Many more photon-neutrino processes become possible
if one admits neutrino masses or
anomalous magnetic dipole moments
\cite{ioaraf}.

In astrophysical environments it is often not realistic
to consider these processes as occuring in vacuum. Plasma
effects must be taken into account, 
as well as the presence of magnetic fields
which, at the surface of neutron stars,
have recently been found to surpass the ``critical'' magnetic field
strength $B_{\rm crit} = {m_e^2\over e} =
4.41\times 10^{13}$ Gauss
\cite{kaspi}.
Of particular interest are then processes
which 
do not occur in vacuum but become possible
in a medium or B-field. An important example is
the plasmon decay $\gamma\to\nu\bar\nu$
\cite{adruwo,zaidi}, 
believed to be an important source for
neutrino production in some types of stars \cite{raffelt}.
Similarly the Cherenkov process $\nu\to\nu\gamma$
becomes possible, although it turns out to be of 
lesser astrophysical relevance \cite{galnik,ioaraf}.
For processes of this type the magnetic field plays a
double role. Firstly, it provides an effective
photon-neutrino coupling via intermediate
charged particles \cite{galnik,skobelev,demida}. 
Secondly, by modifying the photon dispersion relations
it opens up phase space for neutrino-photon reactions
of the type $1\to 2 + 3$.

Similarly one would expect the magnetic field 
to remove the Fermi limit
suppression of the above $2\to 2$ processes.
This has recently been verified both for the
weak \cite{shaisultanov97,chklnt} and 
strong field cases \cite{chkumi}.

In the standard model
the effective coupling is provided by the diagrams
shown in fig. 2(a) and 2(b).
The double line represents the
electron propagator in the presence of the B-field
\footnote{I thank A.N. Ioannisian for providing this
figure.}
.

\begin{figure}
\centering
%\leavevmode
\vbox{
\unitlength=0.8mm
\begin{picture}(60,25)
\put(8,15){\line(-1,1){8}}
\put(8,15){\line(-1,-1){8}}
\put(0,7){\vector(1,1){4}}
\put(8,15){\vector(-1,1){6}}
\multiput(9.5,15)(6,0){3}{\oval(3,3)[t]}
\multiput(12.5,15)(6,0){3}{\oval(3,3)[b]}
\put(31,15){\circle{10}}
\put(31,15){\circle{9}}
\multiput(37.5,15)(6,0){3}{\oval(3,3)[t]}
\multiput(40.5,15)(6,0){3}{\oval(3,3)[b]}
\put(0,10){\shortstack{{}$\nu$}}
\put(18,18){\shortstack{{$Z$}}}
\put(43,18){\shortstack{{$\gamma$}}}
\put(30,11){\shortstack{{e}}}
\put(57,13){\shortstack{{(a)}}}
\end{picture}

\unitlength=0.8mm
\begin{picture}(60,32)
\put(16,15){\line(-1,1){7.5}}
\put(16,15){\line(-1,-1){7.5}}
\put(15,15){\line(-1,1){7}}
\put(15,15){\line(-1,-1){7}}
\put(16,15){\line(-1,1){16}}
\put(16,15){\line(-1,-1){16}}
\put(1,0){\vector(1,1){6}}
\put(16,15){\vector(-1,1){10}}
\multiput(17.5,15)(6,0){3}{\oval(3,3)[t]}
\multiput(20.5,15)(6,0){3}{\oval(3,3)[b]}
\multiput(8.2,8.7)(0,6){3}{\oval(3,3)[l]}
\multiput(8.2,11.7)(0,6){2}{\oval(3,3)[r]}
\put(2,15){\shortstack{{}$W$}}
\put(0,3){\shortstack{{}$\nu$}}
\put(23,18){\shortstack{{$\gamma$}}}
\put(57,13){\shortstack{{(b)}}}
\end{picture}

\vskip3ex

\unitlength=0.8mm
\begin{picture}(60,25)
\put(8,15){\line(-1,1){8}}
\put(8,15){\line(-1,-1){8}}
\put(0,7){\vector(1,1){4}}
\put(8,15){\vector(-1,1){6}}
\put(13,15){\circle{10}}
\put(13,15){\circle{9}}
\multiput(19.5,15)(6,0){3}{\oval(3,3)[t]}
\multiput(22.5,15)(6,0){3}{\oval(3,3)[b]}
\put(0,10){\shortstack{{}$\nu$}}
\put(26,18){\shortstack{{$\gamma$}}}
\put(12,11){\shortstack{{e}}}
\put(57,13){\shortstack{{(c)}}}
\end{picture}
}
\smallskip
\smallskip
\caption[...]{Neutrino-photon coupling in an external magnetic field.
(a)~$Z$-$A$-mixing. (b)~Penguin diagram (only for $\nu_e$).
(c)~Effective coupling in the Fermi limit.
\label{Fig2}}
\end{figure}
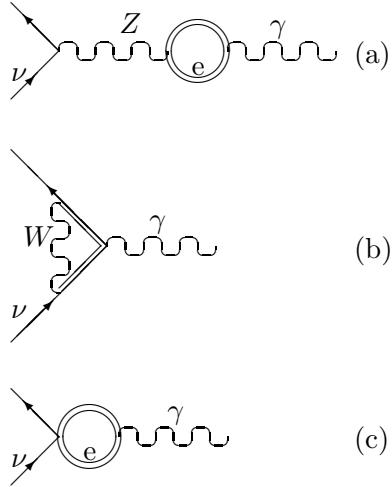

\no
In the limit of infinite gauge-boson masses
both diagrams can be replaced by diagram fig. 2(c).
The amplitude then effectively reduces
to a photonic amplitude with one of the
photons replaced by the neutrino current.

One is thus led to the study of vector -- axialvector
amplitudes in a constant field. In the string-inspired
formalism, various different representations have been
derived for axialvector couplings \cite{mnss,dhogag,mcksch}.
We will use here the one proposed in \cite{mcksch}
and elaborated in \cite{dimcsc}, which has the advantage of
avoiding the separation of these amplitudes into
their real and imaginary parts. 

Extending that work to the case where an
additional constant electromagnetic background
field is present will enable us to perform the
first calculation of the vector -- axialvector amplitude
in a general such field. This amplitude is of relevance
for various of the above processes. It has been obtained 
previously for the magnetic 
\cite{demida,ioaraf,bokumi}
and crossed field \cite{galnik,bokumi} special cases.

The organization of the paper is the following. In the second
chapter, we shortly review the worldline representation of
vector - axialvector amplitudes proposed in \cite{mcksch,dimcsc}.
We then extend this formalism to the inclusion of constant external
fields in chapter three. Chapter four contains a detailed
calculation of the vector - axialvector amplitude in a constant
field. The result is discussed in chapter five, as well as
possible generalizations.

%\vfill\eject
\vspace{30pt}
\section{
Worldline Representation of the Vector-Axialvector Effective
Action
}
\renewcommand{\theequation}{2.\arabic{equation}}
\setcounter{equation}{0}

In \cite{mcksch} the
following first-quantized path integral representation
was derived for the 
one-loop 
effective action induced by a Dirac
fermion loop for a background 
vector field $A$ and axial vector field 
$A_5$ (both abelian)
\footnote{%
We work initially in the Euclidean with
a positive definite metric
$g_{\mu\nu}={\,\mathrm diag}(++++)$.
The Euclidean field strength tensor is defined by
$F^{ij}= \varepsilon_{ijk}B_k, i,j = 1,2,3$,
$F^{4i}=-iE_i$, its dual by
$\tilde F^{\mu\nu} = \half 
\varepsilon^{\mu\nu\alpha\beta}F^{\alpha\beta}$
with $\varepsilon^{1234} = 1$.  
To obtain the corresponding Minkowski space amplitudes
replace
$g_{\mu\nu}\rightarrow \eta_{\mu\nu}
= {\,\mathrm diag}(-+++)$, $
k^4\rightarrow -ik^0, T\rightarrow is,
\varepsilon^{1234}
\rightarrow
i\varepsilon^{1230},
\varepsilon^{0123}=1,
F^{4i}\rightarrow F^{0i}=E_i,
\tilde F^{\mu\nu}\rightarrow -i\tilde F^{\mu\nu}$.
},

\bear
\Gamma[A,A_5]&=&
\ln {\rm Det}
[\slash p +e\slash A
+e_5\gamma_5{\slash A}_5
-im 
]\non\\
&=&
-\half
\int_0^{\infty}{dT\over T}
\,\e^{-m^2T}
\Dx
\int
{\cal D}\psi
\,\,\e^{-\int_0^Td\tau\, L(\tau)}\\
L &=& 
\kinb + \half\psi\cdot\dot\psi + ie\dot x\cdot A 
-ie\psi\cdot F\cdot\psi
\non\\
&& +ie_5\hat\gamma_5
\Bigl(
-2\dot x\cdot\psi\psi\cdot A_5
+\partial\cdot A_5
\Bigr)
+ (D-2)e_5^2 A_5^2
\non\\
\label{GammaAA5}
\ear\no
Here $T$ denotes the usual Schwinger proper-time for the
loop fermion,
$\int {\cal D}x$ the integral over the space of all closed
loops in 
spacetime with periodicity $T$, and 
$\int {\cal D}\psi$ a Grassmann path integral representing the
spin of the loop fermion.
The boundary conditions on the Grassmann path integral are,
after expansion of the interaction exponential, determined by the
power of $\hat\gamma_5$ appearing in a given term; they are
periodic (antiperiodic) with period $T$ if that power is odd (even).
After the boundary conditions are determined $\hat\gamma_5$ can be
replaced by unity.

Note that, although we will consider only the four-dimensional
case in the present work, the Lagrangian has an explicit
dependence
on the spacetime dimension $D$ through
dimensional regularisation \cite{dimcsc}.

Before evaluating this double path integral
one must eliminate the zero mode(s)
contained in it. 
For the $N$ - photon amplitude, 
the Grassmann path integral has to be taken with
antiperiodic boundary conditions, so that only 
$\int{\cal D}x$ has a zero mode.
This zero mode we eliminated by fixing the
average position
$
x_0^{\mu}\equiv {1\over T}\int_0^T d\tau\, x^{\mu}(\tau)
$
of the loop,

\bear
x^{\mu}(\tau) &=& x_0^{\mu} + y^{\mu}(\tau)\non\\
\Dx &=& \int dx_0 \Dy\non\\
\label{split}
\ear\no
When appearing with periodic boundary conditions,
the Grassmann path integral has a zero mode as well.
This can be removed analogously,

\bear
\psi^{\mu}(\tau) &=& \psi_0^{\mu} + \xi^{\mu}(\tau)
\label{splitgrass}\non\\
\int_0^Td\tau \, \xi^{\mu}(\tau) &=& 0
\non\\\label{xicond}
\ear\no
The zero mode integration then produces the
$\varepsilon$ - tensor expected for a fermion loop
with an odd number of axial insertions,

\be
\int d^4\psi_0
\psi_0^{\mu}\psi^{\nu}_0\psi^{\kappa}_0\psi^{\lambda}_0
=\varepsilon^{\mu\nu\kappa\lambda}
\label{zeromodeintegral}
\ee\no
The reduced path integrals will, following 
the recipes of the ``string-inspired''
formalism, be evaluated using Green's
functions in $\tau$ - space. 
For the case of an even number of axial vectors
the Wick contraction rules are the same as in the
QED case treated in part I,

\begin{eqnarray}
\langle y^\mu(\tau_1)\,y^\nu(\tau_2)\rangle
&=& -g^{\mu\nu}{G}_{B}(\tau_1,\tau_2)\label{corry}\\
\langle\psi^{\mu}(\tau_1)\, \psi^{\nu}(\tau_2)\rangle
&=& 
\half g^{\mu\nu}{G}_{F}(\tau_1,\tau_2)\label{corrpsi}
\end{eqnarray}\no
where 

\begin{eqnarray}
G_B(\tau_1,\tau_2)
    &=& \mid \tau_1 - \tau_2\mid -
{{(\tau_1 - \tau_2)}^2\over T}\non\\
G_F(\tau_1,\tau_2)
    &=& {\rm sign} (\tau_1-\tau_2)\non\\
\label{defGBGF}
\end{eqnarray}
\no
In the case of an odd number of axialvectors the
Grassmann Wick contraction rule (\ref{corrpsi})
has to be replaced by 

\bear
\langle\xi^{\mu}(\tau_1)\, \xi^{\nu}(\tau_2)\rangle
&=& 
\half g^{\mu\nu}{\dot{G}}_{B}(\tau_1,\tau_2)
\label{corrxi}
\end{eqnarray}\no
(A ``dot'' always refers to a derivative in the
first variable.)
The 
free Gaussian path integral
determinants are, in our conventions,

\bear
\int{\cal D}y
\,\e^{-\int_0^T d\tau\, \fourth
\dot y^2}
&=& {(4\pi T)}^{-{D\over 2}} \label{freepiy}\\
\int_A{\cal D}\psi\, \e^{-\int_0^Td\tau \,\half\psi\cdot\dot\psi}
&=& 4\label{freepipsi}\\
\int_P{\cal D}\xi\, \e^{-\int_0^Td\tau \,\half\xi\cdot\dot\xi}
&=& 1\label{freepixi}
\ear
In writing them we have specialized to
the four-dimensional case, leaving a $D$ - dependence only 
for the free $y$ - path integral (\ref{freepiy}). This 
anticipates the dimensional regularization.
See \cite{dimcsc} for the case of an arbitrary (even) spacetime
dimension.

\section{
One-Loop Vector-Axialvector Amplitudes in a Constant Field
}
\renewcommand{\theequation}{3.\arabic{equation}}
\setcounter{equation}{0}

In the same way as it was done in part I for the pure vector case,
we can obtain from (\ref{GammaAA5}) the
$M$ vector -- $N$ axialvector amplitude
by specializing to backgrounds
consisting of 
sums of plane waves with definite
polarizations,

\bear
A_{\mu}(x)&=&
\sum_{i=1}^M
\varepsilon_{i\mu}
\e^{ik_i\cdot x}\non\\
A_{5\mu}(x)&=&
\sum_{i=1}^N
\varepsilon_{5i\mu}
\e^{ik_{5i}\cdot x}\non\\
\label{planewavebackground}
\ear\no
and picking out the term containing every
$\varepsilon_i,\varepsilon_{5i}$ once.
The only slight complication compared to the
QED case is due to the term quadratic in $A_5$
appearing in the worldline
Lagrangian (\ref{GammaAA5}). 
To be able to define an analogue of the photon
vertex operator

\begin{equation}
V_A^{\half}[k,\varepsilon]
=
\int_0^Td\tau
\Bigl(
\varepsilon\cdot \dot x
+2i
\varepsilon\cdot\psi
k\cdot\psi
\Bigr)\,
{\rm e}^{ikx}
\label{photonvertopspin}
\end{equation}
\noindent
for the axial coupling, it is convenient to linearize this
term through the introduction
of an auxiliary path
integration \cite{dimcsc}:

\bear
\label{linearizeA52}
\exp \Bigl[-(D-2)e_5^2\int_0^Td\tau A_5^2\Bigr]
&=&
\int {\cal D}z 
\non\\
&&
\hspace{-100pt}\times 
\exp \Bigl[-\int_0^Td\tau 
\Bigl({z^2\over 4}+ie_5\sqrt{D-2}z\cdot A_5\Bigr)
\Bigr]
\non\\
\ear\no
The Wick contraction rule for this auxiliary field is
simply

\bear
\langle z^{\mu}(\tau_1)z^{\nu}(\tau_2)\rangle
&=&
2g^{\mu\nu}\delta (\tau_1-\tau_2)
\label{wickz}
\ear\no
This allows us to define an axial-vector vertex operator
as follows,

\bear
V_{A_5}[k,\varepsilon] &\equiv&
\hat\gamma_5
\int_0^Td\tau
\Bigl(i\varepsilon\cdot k 
+ 2\varepsilon\cdot\psi
\dot x\cdot\psi
+ \sqrt{D-2}\,\varepsilon\cdot z
\Bigr)
\,\e^{ik\cdot x}
\non\\
\label{defaxvectvertop}
\ear\no
We can then write the
one-loop $M$ vector --  $N$ axialvector
amplitude in terms of Wick contractions of
vertex operators:

\bear
\Gamma[\lbrace k_i,\varepsilon_i\rbrace,
\lbrace k_{5j},\varepsilon_{5j}\rbrace]
&=&
-\half N_{A,P}(-i)^{M+N}e^Me_5^N
\non\\&&
\hspace{-95pt}
\times
\Tintm 
\PITD
\Bigl\langle
V_{A}^{\half}[k_1,\varepsilon_1]\ldots
\non\\&&
\hspace{-95pt}
\ldots
V_{A}^{\half}[k_M,\varepsilon_M]
V_{A_5}[k_{51},\varepsilon_{51}]\ldots
V_{A_5}[k_{5N},\varepsilon_{5N}]
\Bigr\rangle_{A,P}
\non\\
\label{repMvectorNaxial}
\ear\no
The Wick contractions are done using (\ref{corry}),
(\ref{wickz}), and either (\ref{corrpsi}) or (\ref{corrxi}),
depending on the boundary conditions.
This can still be done
in closed form \cite{dimcsc},
however the result is lengthy.

Let us now introduce an additional background vector field
$\bar A^{\mu}(x)$
with constant field strength tensor
$\bar F_{\mu\nu}$. 
In Fock--Schwinger gauge
centered at $x_0$ \cite{ss1} 
its contribution to the
worldline Lagrangian (\ref{GammaAA5}) 
can be written as
$
\Delta L = {1\over 2}\,ie\,y^{\mu} \bar F_{\mu\nu}
\dot y^{\nu} - ie\,\psi^{\mu} \bar F_{\mu\nu}\psi^{\nu}
$.
It can therefore be absorbed into the kinetic part of Lagrangian,
and be taken into account
by a change of the
free worldline propagators.
This leads to a replacement of 
$G_B,\dot G_B,G_F$ by (\cite{cadhdu,shaisultanov,rss}; see also
\cite{mckshe})

\bear
{{\cal G}_B}(\tau_1,\tau_2) &=&
{T\over 2{\cal Z}^2}
\biggl({{\cal Z}\over{{\rm sin}({\cal Z})}}
\,{\rm e}^{-i{\cal Z}\dot G_{B12}}
+ i{\cal Z}\dot G_{B12} - 1\biggr)
\non\\
\dot{\cal G}_B(\tau_1,\tau_2) &=&
{i\over {\cal Z}}
\biggl({{\cal Z}\over{{\rm sin}({\cal Z})}}
\,{\rm e}^{-i{\cal Z}\dot G_{B12}}
- 1\biggr)
\non\\
{\cal G}_{F}(\tau_1,\tau_2) &=&
G_{F12} {{\rm e}^{-i{\cal Z}\dot G_{B12}}\over {\rm cos}({\cal Z})}
\non\\
\label{calGBGF}
\ear
where we have defined ${\cal Z}\equiv e{F}T$
(omitting the ``bar''). 
The presence of the background field thus modifies 
the Wick contraction rules 
eqs.(\ref{corry}),(\ref{corrpsi}),(\ref{corrxi})
to

\begin{eqnarray}
\langle y^{\mu}(\tau_1)y^{\nu}(\tau_2)\rangle
&=&
-{\cal G}_B^{\mu\nu}(\tau_1,\tau_2)\label{corryF}\\
\langle\psi^{\mu}(\tau_1)\psi^{\nu}(\tau_2)\rangle
&=&
\frac{1}{2}{\cal G}_F^{\mu\nu}(\tau_1,\tau_2)\label{corrpsiF}\\
\langle\xi^{\mu}(\tau_1)\xi^{\nu}(\tau_2)\rangle
&=&
\frac{1}{2}{\dot{\cal G}}_B^{\mu\nu}(\tau_1,\tau_2)\label{corrxiF}
\end{eqnarray}
\noindent
The free Gaussian path integral determinants 
(\ref{freepiy}),(\ref{freepipsi}),(\ref{freepixi})
also become field dependent \cite{ss1,rss}:

\bear
\int{\cal D}y
\,\e^{-\int_0^T d\tau\, \Bigl( \fourth
\dot y^2 +
{1\over 2} ie\,y^{\mu} F_{\mu\nu}
\dot y^{\nu} \Bigr)
}
&=& {(4\pi T)}^{-{D\over 2}} 
{\rm det}^{-{1\over 2}}
\biggl[{\sin({\cal Z})\over {{\cal Z}}}
\biggr]
\label{freepiyF}\\
\int_A{\cal D}\psi\, \e^{-\int_0^Td\tau 
\Bigl(
\half\psi\cdot\dot\psi
- ie\,\psi^{\mu} F_{\mu\nu}\psi^{\nu}
\Bigr)
}
&=& 4\,
{\rm det}^{\half}\Bigl[\cos{\cal Z}\Bigr]
\label{freepipsiF}\\
\int_P{\cal D}\xi\, \e^{-\int_0^Td\tau 
\Bigl( \half\xi\cdot\dot\xi
- ie\,\xi^{\mu} F_{\mu\nu}\xi^{\nu}
\Bigr)
}
&=& 
{\rm det}^{1\over 2}
\biggl[{\sin({\cal Z})\over {{\cal Z}}}
\biggr]
\label{freepixiF}
\ear\no
Note, however, that for the case of periodic
Grassmann boundary conditions this field dependence
cancels out between the coordinate and Grassmann
path integrals. This cancellation can be understood
as a consequence of the fact that the two
path integrals are related by worldline supersymmetry
\cite{berkos,strassler,dimcsc}. It does not occur in the
antiperiodic case since here the supersymmetry is broken
by the boundary conditions.

The above trigonometric expressions should be understood as power
series in the field strength tensor. 
In section 3.2 of part I we gave explicit expressions for the
generalized worldline Green's functions and determinants
in terms of $\Eins, F, \tilde F, F^2$, and the two
standard Maxwell invariants 
$f=\fourth F\cdot F, 
g= \fourth F\cdot\tilde F$.

This is all, then, which we have to know to calculate 
one-loop processes involving any numbers of (abelian)
vectors, axialvectors as well as a constant external (vector)
field. While in the present paper we will consider only
the amplitude case, the formalism can be applied as well to the
calculation of the effective action itself in the inverse
higher derivative expansion 
(see \cite{ss1,fhss,cadhdu,gussho,shovkovy,mcksch}).

\vspace{30pt}
\section{Worldline Calculation of
the Vector -- Axialvector Amplitude in a Constant Field}
\renewcommand{\theequation}{4.\arabic{equation}}
\setcounter{equation}{0}

As an explicit example, we calculate the
vector -- axialvector two-point function
in a constant field.
According to the above we can represent this amplitude as follows,

\bear
\langle
A_{\mu}(k_1)
A_{5\nu}(k_2)
\rangle
&=&
{1\over 2}
\Tintm
\Dx
\int{\cal D}\psi
\non\\
&&\!\!\!\!\!\!\!\!\!\!\!\!\!\!\!\!
\times\exp\biggl\lbrace
-\int_0^Td\tau
\Bigl[
\kinb
+{1\over 2}
\psi\cdot\dot\psi
+{i\over 2}e\, x\cdot F\cdot\dot x
-ie\, \psi\cdot F\cdot\psi
\Bigr]
\biggr\rbrace\non\\
&&\!\!\!\!\!\!\!\!\!\!\!\!\!\!\!\!
\times
\int_0^Td\tau_1
\Bigl(
\dot x_{\mu}(\tau_1)+2i\psi_{\mu}(\tau_1)
k_1\cdot\psi(\tau_1)
\Bigr)
\e^{ik_1\cdot x_1}
\non\\&&\!\!\!\!\!\!\!\!\!\!\!\!\!\!\!\!\times
\int_0^Td\tau_2
\Bigl(
ik_{2\nu}+2\psi_{\nu}(\tau_2)\dot x(\tau_2)\cdot\psi(\tau_2)
\Bigr)
\e^{ik_2\cdot x_2}
\non\\
\label{AA5withF}
\ear
Note that this expression is already manifestly gauge
invariant, i.e. transversal in the vector index.
If one multiplies the right hand side by $k_1^{\mu}$
then the integrand of the vector vertex operator
becomes a total derivative in $\tau_1$, 
so that the integral vanishes by
periodicity. This mechanism is
well-known from string theory. 
Nothing analogous holds for the
axialvector vertex operator.

This amplitude is finite, so that we can set $D=4$ in
its evaluation.
As a first step, the zero-modes of both path integrals
are separated out according to 
eqs.(\ref{split}),(\ref{splitgrass}),
and the Grassmann zero mode integrated out using 
eq.(\ref{zeromodeintegral}).
All terms which do not contain all four zero mode components
precisely once give zero. To explicitly perform this
integration we
note that by eq.(\ref{xicond}) we can rewrite, in the exponent
of eq.(\ref{AA5withF}),

\be
\int_0^Td\tau\, \psi(\tau)\cdot F\cdot\psi(\tau)
=
T
\psi_0\cdot F\cdot\psi_0
+
\int_0^Td\tau\,
\xi(\tau)\cdot F\cdot\xi(\tau)
\label{splitpsiFpsi}
\ee\no
Thus for 
the case at hand the Grassmann zero mode integral can appear
in the following three forms,

\bear
\int d^4\psi_0 
\,\e^{ieT\psi_0\cdot F\cdot\psi_0}
&=&
-{(eT)^2\over 2}
\varepsilon_{\mu\nu\kappa\lambda}
F_{\mu\nu}F_{\kappa\lambda}
=
-(eT)^2 F\cdot\tilde F\non\\
\int d^4\psi_0 
\,\e^{ieT\psi_0\cdot F\cdot\psi_0}
\psi_{0\mu}\psi_{0\nu}
&=&
ieT\varepsilon_{\mu\nu\kappa\lambda}F_{\kappa\lambda}
= 2ieT\tilde F_{\mu\nu}
\non\\
\int d^4\psi_0
\,\e^{ieT\psi_0\cdot F\cdot\psi_0}
\psi_{0\mu}\psi_{0\nu}\psi_{0\kappa}\psi_{0\lambda} 
&=&
\varepsilon_{\mu\nu\kappa\lambda}
\non\\
\label{zeromodeintegrals}
\ear\no
%Here and in the following we again absorb a factor of
%$eT$ into $F,\tilde F$.
In the next step, both path integrations are performed
using the field-dependent Wick contraction rules
eqs. (\ref{corryF}) and (\ref{corrxiF}). 
This results in the following parameter integral 
representation for the vector -- axialvector
vacuum polarisation tensor
\footnote{
Since $G_F,{\cal G}_F$ do not occur for the periodic
case we delete the subscript ``B'' in the following.}

\bear
\Pi_5^{\mu\nu}(k)
&=&
{e e_5\over 8\pi^2}
\int_0^{\infty}{dT\over T^3}\,\e^{-m^2T}
\int_0^Td\tau_1 d\tau_2
\,\,
J_5^{\mu\nu}(\tau_1,\tau_2)
\e^{-k\cdot \bar{\cal G}_{12}\cdot k}\non\\
J_5^{\mu\nu}(\tau_1,\tau_2)&=&
\Bigl[ 
\ddot{\cal G}_{12}^{\mu\alpha}
-
(
\dot{\cal G}_{21}^{\alpha\beta} 
-
\dot{\cal G}_{22}^{\alpha\beta} 
)
(
\dot{\cal G}_{11}^{\mu\rho} 
-
\dot{\cal G}_{12}^{\mu\rho} 
)
k_{\beta}k_{\rho}
\Bigr]
\Bigl(
i\tilde{\cal Z}_{\nu\alpha}
-{{\cal Z}\cdot\tilde{\cal Z}\over 4}
\dot{\cal G}_{22}^{\nu\alpha}
\Bigr)
\non\\
&&\hspace{-65pt}
+
{{\cal Z}\cdot\tilde{\cal Z}\over 4}
(
\dot{\cal G}_{11}^{\mu\rho} 
-
\dot{\cal G}_{12}^{\mu\rho} 
)
k_{\rho}k_{\nu}
+k_{\nu}k_{\rho}
\Bigl(
i\tilde{\cal Z}_{\mu\rho}
- {{\cal Z}\cdot\tilde{\cal Z}\over 4}
\dot{\cal G}_{11}^{\mu\rho}
\Bigr)
+k_{\rho}k_{\sigma}
(
\dot{\cal G}_{21}^{\alpha\rho}
-
\dot{\cal G}_{22}^{\alpha\rho}
)
\non\\
&&\hspace{-65pt}\times
\Bigl[
\varepsilon_{\mu\sigma\nu\alpha}
+i\bigl(
\dot{\cal G}_{22}^{\nu\alpha}
\tilde{\cal Z}_{\mu\sigma}
-
\dot{\cal G}_{12}^{\sigma\alpha}
\tilde{\cal Z}_{\mu\nu}
+
\dot{\cal G}_{12}^{\sigma\nu}
\tilde{\cal Z}_{\mu\alpha}
+
\dot{\cal G}_{12}^{\mu\alpha}
\tilde{\cal Z}_{\sigma\nu}
-
\dot{\cal G}_{12}^{\mu\nu}
\tilde{\cal Z}_{\sigma\alpha}
+
\dot{\cal G}_{11}^{\mu\sigma}
\tilde{\cal Z}_{\nu\alpha}
\bigr)
\non\\
&&\hspace{-65pt}
- {{\cal Z}\cdot\tilde{\cal Z}\over 4}
\bigl(
\dot{\cal G}_{11}^{\mu\sigma}\dot{\cal G}_{22}^{\nu\alpha}
-\dot{\cal G}_{12}^{\mu\nu}\dot{\cal G}_{12}^{\sigma\alpha}
+\dot{\cal G}_{12}^{\mu\alpha}\dot{\cal G}_{12}^{\sigma\nu}
\bigr)
\Bigr]
\non\\
\label{Pi5}
\ear\no
where
$k=k_1=-k_2, \tilde {\cal Z} \equiv eT\tilde F,
\bar {\cal G}_{12} 
\equiv {\cal G} (\tau_1,\tau_2)
-{\cal G} (\tau,\tau)$
(see part I). 
As in the vector -- vector case, it is useful
to perform a partial integration on the one term 
involving $\ddot{\cal G}_{12}$, leading to the replacement

\be
\ddot{\cal G}_{12}^{\mu\alpha}
\rightarrow
\dot{\cal G}_{12}^{\mu\alpha}
k\cdot\dot{\cal G}_{12}\cdot k
\label{partintddotG}
\ee\no
By this partial integration, and the removal of some
terms which cancel against each other,
$J_5^{\mu\nu}(\tau_1,\tau_2)$ gets replaced 
by 

\bear
&&\hspace{-20pt}
k^{\rho}k^{\sigma}
\Bigl[
\dot {\cal G}^{\mu\alpha}_{12}
\dot{\cal G}_{12}^{\rho\sigma}
+
(
\dot{\cal G}_{21}^{\alpha\sigma} 
-
\dot{\cal G}_{22}^{\alpha\sigma} 
)
\dot{\cal G}_{12}^{\mu\rho} 
\Bigr]
\Bigl(
i\tilde{\cal Z}^{\nu\alpha}
-{{\cal Z}\cdot\tilde{\cal Z}\over 4}
\dot{\cal G}_{22}^{\nu\alpha}
\Bigr)
-
{{\cal Z}\cdot\tilde{\cal Z}\over 4}\dot{\cal G}_{12}^{\mu\rho} 
k^{\rho}k^{\nu}
\non\\
&&\hspace{-20pt}
+ik^{\nu}k^{\rho}
\tilde{\cal Z}^{\mu\rho}
+k^{\rho}k^{\sigma}
(
\dot{\cal G}_{21}^{\alpha\rho}
-
\dot{\cal G}_{22}^{\alpha\rho}
)
\Bigl[
\varepsilon^{\mu\sigma\nu\alpha} 
+ {{\cal Z}\cdot\tilde{\cal Z}\over 4}
\bigl(
\dot{\cal G}_{12}^{\mu\nu}\dot{\cal G}_{12}^{\sigma\alpha}
-\dot{\cal G}_{12}^{\mu\alpha}\dot{\cal G}_{12}^{\sigma\nu}
\bigr)
\non\\
&&\hspace{-20pt}
+i\bigl(
\dot{\cal G}_{22}^{\nu\alpha}
\tilde{\cal Z}^{\mu\sigma}
-
\dot{\cal G}_{12}^{\sigma\alpha}
\tilde{\cal Z}^{\mu\nu}
+
\dot{\cal G}_{12}^{\sigma\nu}
\tilde{\cal Z}^{\mu\alpha}
+
\dot{\cal G}_{12}^{\mu\alpha}
\tilde{\cal Z}^{\sigma\nu}
-
\dot{\cal G}_{12}^{\mu\nu}
\tilde{\cal Z}^{\sigma\alpha}
\bigr)
\Bigr]\non\\
\label{J5munusimpler}
\ear\no
Next we decompose ${\cal G}_{ij}$ 
as

\bear
{\cal G}_{ij}
&=&
{\cal S}_{ij}
+
{\cal A}_{ij}
\label{decomposecalGBGF}
\ear\no
where ${\cal S}$ (${\cal A}$) are its parts
even (odd) in $F$. 
We can then delete all terms
odd in $\tau_1 -\tau_2$ since they vanish
upon integration.
After using the identity $F\tilde F = -g\Eins$
and some combining of terms, 
$J_5$ finally turns into the following,
nicely symmetric expression $I_5$,

\bear
I_5^{\mu\nu}(\tau_1,\tau_2) &=&
i\biggl\lbrace
\tilde{\cal Z}^{\mu\nu}k{\cal U}_{12}k
+\Bigl[
(\tilde{\cal Z} k)^{\mu}({\cal U}_{12}k)^{\nu}
+ (\mu\leftrightarrow\nu )
\Bigr]
\non\\
&&\quad
-(\tilde{\cal Z}{\dot{\cal S}}_{12})^{\mu\nu}k{\dot{\cal S}}_{12}k
-\Bigl[
(\tilde{\cal Z} {\dot{\cal S}}_{12}k)^{\mu}({\dot{\cal S}}_{12}k)^{\nu}
+ (\mu\leftrightarrow\nu )
\Bigr]
\biggr\rbrace
\non\\
&&\hspace{-20pt}
+{{\cal Z}\cdot\tilde{\cal Z}\over 4}
\biggl\lbrace
-{\dot{\cal A}}_{12}^{\mu\nu}k{\cal U}_{12}k
-\Bigl[
({\dot{\cal A}}_{12}k)^{\mu}({\cal U}_{12}k)^{\nu}
+ (\mu\leftrightarrow\nu )
\Bigr]
\non\\
&&\quad
+({\dot{\cal A}}_{22}
{\dot{\cal S}}_{12}
)^{\mu\nu}
k{\dot{\cal S}}_{12}k
+\Bigl[
({\dot{\cal A}}_{22}{\dot{\cal S}}_{12}k)^{\mu}
({\dot{\cal S}}_{12}k)^{\nu}
+ (\mu\leftrightarrow\nu)
\Bigr]
\biggr\rbrace
\non\\
\label{defI5munu}
\ear\no
Here in addition to ${{\cal A}}$ and ${{\cal S}}$ 
we have introduced the combination
${\cal U}$,

\bear
{\cal U}_{12} &=& {\dot{\cal S}}_{12}^2 - ({\dot{\cal A}}_{12}
-{\dot{\cal A}}_{22})\bigl({\dot{\cal A}}_{12}+{i\over {\cal Z}}\bigr)
   =  {1-\cos({\cal Z}\dot G_{12})\cos({\cal Z})
   \over \sin^2({\cal Z})} \non\\
\label{U}
\ear\no
Defining also

\bear
\hat{\cal A} \equiv \dot{\cal A} + {i\over {\cal Z}}
\label{defhatA}
\ear\no
this expression can be further compressed to

\bear
I_5^{\mu\nu}(\tau_1,\tau_2) &=&
{{\cal Z}\cdot\tilde{\cal Z}\over 4}
\biggl\lbrace
-{\hat{\cal A}}_{12}^{\mu\nu}k{\cal U}_{12}k
-\Bigl[
({\hat{\cal A}}_{12}k)^{\mu}({\cal U}_{12}k)^{\nu}
+ (\mu\leftrightarrow\nu )
\Bigr]
\non\\
&&\quad
+({\hat{\cal A}}_{22}
{\dot{\cal S}}_{12}
)^{\mu\nu}
k{\dot{\cal S}}_{12}k
+\Bigl[
({\hat{\cal A}}_{22}{\dot{\cal S}}_{12}k)^{\mu}
({\dot{\cal S}}_{12}k)^{\nu}
+ (\mu\leftrightarrow\nu)
\Bigr]
\biggr\rbrace
\non\\
\label{defI5munusimp}
\ear\no
We can now use the matrix decompositions of
${\cal S},\dot{\cal S},\dot{\cal A}$, given 
in eq.(3.29) of part I, to write the integrand
in explicit form. In this we have a choice between
the matrix bases $\lbrace\hat{\cal Z}_{\pm},
\hat{\cal Z}_{\pm}^2\rbrace$ 
or 
$\lbrace\Eins, F,\tilde F, F^2\rbrace$. We will use the former
one here since it leads to a somewhat more succinct
expression.
After the usual rescaling to the unit circle,
a transformation of variables $v=\dot G_{12}$, and
continuation to Minkowski
space
\footnote{For the Maxwell invariants this means
$f\rightarrow {\cal F}$, 
$g\rightarrow i{\cal G}$
(we assume 
${\cal G} \geq 0$).}
, we obtain our final result for the vector --
axialvector amplitude in a constant field
\footnote{In undecomposed form this result was already presented in
\cite{mecorfu}.}
,

\bear
\Pi_5^{\mu\nu}(k)
&=&
{e^3e_5\over 8\pi^2}
{\cal G}
\int_0^{\infty}
ds\,s\,\e^{-ism^2}
\int_{-1}^1{dv\over 2}
\,{\rm exp}\biggl[
-i{s\over 2}\sum_{\alpha=+,-}
{\hat A_{B12}^{\alpha}
-\hat A_{B11}^{\alpha}\over z_{\alpha}}\,
k\cdot \hat{\cal Z}_{\alpha}^2\cdot k
\biggr]
\non\\&&\hspace{-15pt}\times
\, \sum_{\alpha,\beta=+,-}
\biggl[
\hat A_{12}^{\alpha}
\Bigl(
(\hat A_{12}^{\beta}-\hat A_{22}^{\beta})
\hat A_{12}^{\beta}
-
(S_{12}^{\beta})^2
\Bigr)
+\hat A_{22}^{\alpha}
S_{12}^{\alpha}S_{12}^{\beta}
\biggr]
\non\\&&\hspace{25pt}\times
\Bigl[\hat{\cal Z}_{\alpha}^{\mu\nu}k\hat{\cal Z}_{\beta}^2k
+(\hat{\cal Z}_{\alpha}k)^{\mu}(\hat{\cal Z}_{\beta}^2k)^{\nu}
+(\hat{\cal Z}_{\alpha}k)^{\nu}(\hat{\cal Z}_{\beta}^2k)^{\mu}
\Bigr]
\non\\
\label{axvpfinal}
\ear
where 

\bear
S_{12}^{\pm} &=&
{\sinh(z_{\pm}\dot G_{B12})\over \sinh(z_{\pm})} 
\non\\
\hat A_{12}^{\pm} &=&
{\cosh(z_{\pm} \dot G_{B12})\over 
\sinh(z_{\pm})},\quad
\hat A_{ii}^{\pm} =
\coth(z_{\pm})
\non\\
z_+ &=& iesa,\qquad
a = \sqrt{\sqrt{{\cal F}^2+{\cal G}^2}+{\cal F}} 
\non\\
z_- &=& -esb,\qquad
b = \sqrt{\sqrt{{\cal F}^2+{\cal G}^2}-{\cal F}} 
\non\\
\hat{\cal Z}_+ &=& {aF-b\tilde F\over a^2+b^2}\non\\
\hat{\cal Z}_- &=& -i{bF+a\tilde F\over a^2 +b^2} \non\\
\label{defsmink}
\ear\no

As in the vector -- vector case, 
this expression becomes somewhat more
transparent if one specializes to the Lorentz
system where $\bf E$ and $\bf B$ 
are both pointing along the positive z - axis,
${\bf E} = (0,0,E), {\bf B} = (0,0,B)$.
Here one obtains
\vfill\eject
\bear
\Pi_5^{\mu\nu}(k)
&=&i
{e^2e_5\over 8\pi^2}
\int_0^{\infty}
ds
\int_{-1}^1{dv\over 2}\,
\e^{-is\Phi_0}
\sum_{\alpha,\beta = \perp,\parallel}
c^{\alpha\beta}
\Bigl[
\tilde F_{\alpha}^{\mu\nu}k_{\beta}^2
+ (\tilde F_{\alpha}k)^{\mu}k_{\beta}^{\nu}
+ (\tilde F_{\alpha}k)^{\nu}k_{\beta}^{\mu}
\Bigr]
\non\\
\label{axvpfinalspecial}
\ear\no
where $z=eBs, z'=eEs$, $k_{\perp}=(0,k^1,k^2,0)$, 
$k_{\parallel} = (k^0,0,0,k^3)$,

\begin{equation}
(\tilde F_{\parallel})^{\mu\nu} \equiv
\left(
\begin{array}{*{4}{c}}
0&0&0&B\\
0&0&0&0\\
0&0&0&0\\
-B&0&0&0
\end{array}
\right),\qquad
(\tilde F_{\perp})^{\mu\nu} \equiv
\left(
\begin{array}{*{4}{c}}
0&0&0&0\\
0&0&-E&0\\
0&E&0&0\\
0&0&0&0
\end{array}
\right)\nonumber\\
\label{deftildeFperpparallel}\nonumber
\vspace{7 mm}
\end{equation}

\bear
\Phi_0 = m^2 +{k_{\perp}^2\over 2}{\cos(zv)-\cos(z)\over z\sin(z)}
-{k_{\parallel}^2\over 2}{\cosh(z'v)-\cosh(z')\over z'\sinh(z')}
\non\\
\label{Phi0}
\ear

\bear
c^{\perp\perp}
&=&
z{\cos(zv)-\cos(z)\over\sin^3(z)}
\non\\
c^{\perp\parallel}
&=&
{z\cos(zv)\over\sin(z)}
{\cosh(z')\cosh(z'v)-1\over\sinh^2(z')}
-
{z\cos(z)\sin(zv)\over\sin^2(z)}
{\sinh(z'v)\over\sinh(z')}
\non\\
c^{\parallel\perp}
&=&
-{z'\cosh(z'v)\over\sinh(z')}
{\cos(zv)\cos(z)-1\over\sin^2(z)}
-{z'\cosh(z')\sinh(z'v)\over\sinh^2(z')}
{\sin(zv)\over\sin(z)}
\non\\
c^{\parallel\parallel}
&=&
-z'{\cosh(z'v)-\cosh(z')\over\sinh^3(z')}
\non\\
\label{calphabeta}
\ear\no
\vfill\eject
\no
This result can still be slightly simplified using the
relations

\be
\tilde F^{\mu\nu}_{\alpha} k_{\alpha}^2 = 
(\tilde F_{\alpha}k)^{\mu}
k_{\alpha}^{\nu}
-
(\tilde F_{\alpha}k)^{\nu}
k_{\alpha}^{\mu}
\label{perpparallelidentity}
\ee\no
$(\alpha = \perp, \parallel)$.

By taking the limit $z'\rightarrow 0$ in 
(\ref{axvpfinalspecial})
we reproduce the known
result for the vector -- axialvector 
vacuum polarisation tensor in a
constant magnetic field,

\bear
\Pi_5^{\mu\nu}(k)
&=&
{e^2e_5\over 16\pi^2m^2}
\biggl\lbrace
C_{\parallel}
\Bigl[
\tilde F^{\mu\nu}k_{\parallel}^2
+(\tilde F k)^{\mu}k_{\parallel}^{\nu}
+(\tilde F k)^{\nu}k_{\parallel}^{\mu}
\Bigr]
\non\\
&&
+
C_{\perp}
\Bigl[
\tilde F^{\mu\nu}k_{\perp}^2
+(\tilde F k)^{\mu}k_{\perp}^{\nu}
+(\tilde F k)^{\nu}k_{\perp}^{\mu}
\Bigr]
\biggr\rbrace
\label{Pi5magfin}
\ear\no
where
\footnote{Our definition of 
$C_{\parallel}$ differs by a factor of 2
from the one used in \cite{ioaraf}.}

\bear
C_{\parallel}
&=&
im^2\int_0^{\infty}ds\int_{-1}^1dv\,\e^{-is\phi_0}
\half (1-v^2)
\label{Cparmag}\\
C_{\perp}
&=&
im^2\int_0^{\infty}ds\int_{-1}^1dv\,\e^{-is\phi_0}R
\label{Cperpmag}\\
\phi_0 &=& m^2 + {{1-v^2} \over 4}k_{\parallel}^2
+
{\cos (vz)-\cos(z)\over 2z\sin(z)}
k_{\perp}^2
\label{phi0}\\
R &=& {1-v\sin(z)\sin(vz)-\cos(z)\cos(vz)\over \sin^2(z)}
\label{R}
\ear\no

The magnetic special case was first
obtained in \cite{demida} and more recently recalculated
in \cite{ioaraf}. Our
eq. (\ref{Pi5magfin}) 
can be easily identified with the results given by those
authors making use of  
(\ref{perpparallelidentity}).

In a ``crossed field'', defined by 
${\bf E}\perp{\bf B}, E=B$, both invariants vanish.
This case is of importance since a general
constant field can be well-approximated by
a crossed field at sufficiently high energies
(see, e.g., \cite{bokumi}). In a crossed field
$F^3 =0$, so that the power series 
(\ref{calGBGF}) break off after their quadratic terms
(see the appendix of part I). The final parameter integral
therefore becomes much simpler,
\vfill\eject
\bear
\Pi_5^{\mu\nu}(k) &=&
i{e^2e_5\over 4\pi^2}\int_0^{\infty}ds
\int_0^1 du_1
\e^{-i
\Bigl[
sm^2 + sG_{12}k^2 -{s^3\over 3}G_{12}^2e^2
kF^2k\Bigr]}
\non\\
&&\times
\biggl\lbrace
G_{12}\Bigl[
\tilde F^{\mu\nu}k^2 + (\tilde Fk)^{\mu}k^{\nu}
+ (\tilde Fk)^{\nu}k^{\mu}\Bigr]
\non\\
&&
-s^2e^2
G_{12}^2\Bigl[
\tilde F^{\mu\nu}kF^2k + (\tilde Fk)^{\mu}(F^2k)^{\nu}
+ (\tilde Fk)^{\nu}(F^2k)^{\mu}\Bigr]
\biggr\rbrace 
\non\\
\label{axvpcrossed}
\ear\no
($G_{12}=u_1(1-u_1)$).

Note that our final expressions are manifestly transversal in
the vector index. In a field theory calculation 
this would not be automatically the case,
since this amplitude contains the
ABJ anomalous triangle graphs \cite{adler,beljac}.
As was already shown in \cite{dimcsc}
in the present formalism the anomalous divergence is 
unambiguously fixed to be at the axialvector.

\section{Discussion}
\renewcommand{\theequation}{5.\arabic{equation}}
\setcounter{equation}{0}

We have explained in detail
how the string-inspired technique can be applied
to the computation  of one-loop 
amplitudes involving any numbers of vectors and
axialvectors, as well as a constant electromagnetic
field. Our explicit calculation of the vector -- axialvector
two-point amplitude displayed already some of the usual
features of the string-inspired technique.
In the same way as for the
vector -- vector case, the
Bern-Kosower type partial integration procedure has led 
to a manifestly 
gauge invariance result
automatically,
without the need of performing any subtractions.
However, in contrast to the vector -- vector case we have not
been able to avoid here the explicit evaluation of the
Grassmann path integral. In the V-V case this was possible
through the application of the ``cycle replacement rule'',
for which no generalization has been found yet to the
case of mixed vector -- axialvector amplitudes.

Even in the absence of such a generalization
it would be of interest to extend the general analysis of the
partial integration procedure, performed for the
pure vector case in \cite{mepartint},
to the mixed vector -- axialvector case.
Another useful generalization of our results would be the
extension to the finite temperature case along the lines
of \cite{shovkovy,haasch,sato}.

\vskip15pt
\noindent{\bf Acknowledgements:}
I would like to thank S.L. Adler, 
A.N. Ioannisian, N.V. Mikheev, R. Stora, 
and J.-B. Zuber for enlightening discussions and correspondence.
\vskip10pt
\noindent
{\bf Note added:} The recent e-print \cite{shaisultanov2000}
contains a field theory calculation of the vector-axialvector
amplitude in a constant field. The final result (13) of this
calculation agrees with our (\ref{axvpfinalspecial}).

%\vfill\eject

\end{document}